\documentclass[12pt]{article}
\usepackage{psfig}
\def\be{\begin{equation}}
\def\ee{\end{equation}}
\def\beq{\begin{equation}}
\def\eeq{\end{equation}}
\def\beqar{\begin{eqnarray}}
\def\eeqar{\end{eqnarray}}
\def\barr{\begin{array}}
\def\earr{\end{array}}
\def\lsim{\:\raisebox{-0.5ex}{$\stackrel{\textstyle<}{\sim}$}\:}

\def\and{\qquad {\rm and } \qquad}

\def\slp{p \hspace{-1ex}/}
\def\slk{$k$ \hspace{-1ex}/}

\def\slk{k \hspace{-1ex}/}

\def\ie{ {\it i.e.} }

\def\sbar{ \overline{s} }
\def\thmin{\theta_0}
\def\cmin{\cos \theta_0}

\def\eebar{$e^+e^-~$}
\def\ttbar{$t \overline{t}~$}
\def\eegz{$e^+e^- \to \gamma Z~$}
\def\ggz{$\gamma\gamma Z~$}
\def\gzz{$\gamma ZZ~$}
\begin{document}
\renewcommand{\thefootnote}{\fnsymbol{footnote}}
\begin{flushright}
IISc-CHEP-10/04 \\
TU-729\\
hep-ph/0410084
\end{flushright}
\vskip .3cm

\begin{center}{\Large \bf \boldmath
Transverse beam polarization and 
CP violation in
$e^+e^- \to \gamma Z$ with contact interactions}
\vskip 1cm
{B. Ananthanarayan$^a$, Saurabh D. Rindani$^{b,c}$} 
\vskip .5cm

{\it $^a$Centre for High Energy Physics, 
Indian Institute of Science\\ Bangalore
560 012, India\\~ \\
$^b$High Energy Theory Group,
Department of Physics,
Tohoku University\\
Aoba-ku, Sendai 980-8578, Japan\\ ~ \\
$^c$Theory Group, Physical Research Laboratory\\ 
Navrangpura, Ahmedabad 380 009,
India}\footnote{Permanent address}
\end{center}
\vskip 1cm
\begin{quote}
\centerline{\bf Abstract}
We consider the most general gauge-invariant, chirality-conserving
contact interactions in
the process $e^+e^-\to \gamma Z$, of the type proposed Abraham and Lampe,
in order to explore the possibility of CP violation at future
linear colliders in the presence of polarized beams. We hereby
extend recent work on CP violation due to anomalous triple-gauge boson
vertices.
We isolate combinations of
couplings which are genuinely CP violating, pointing out which of these 
can only be studied with the use of transverse polarization.  
We place constraints on these couplings that could arise
from suitably defined CP-odd asymmetries, 
considering realistic polarization (either longitudinal or transverse) 
of 80\% and 60\% for the
electron and positron beams respectively, and with an integrated
luminosity  $\int dt{\cal L}$ of $500\, {\rm fb}^{-1}$ at a
centre of mass energy of $\sqrt{s}=500~{\rm GeV}$. 
\end{quote}

\newpage
\section{Introduction}\label{intro}

An \eebar linear collider (LC) operating at a centre-of-mass (cm) energy of
several hundred GeV is now a distinct possibility. At such a facility, one would  
like to determine precisely known interactions, and discover or
constrain new interactions. Longitudinal polarization of the $e^+$ and
$e^-$ beams, which is expected to be feasible at such colliders, would be
helpful in reducing background as well as enhancing the sensitivity. 
Spin rotators can be used to convert the longitudinal
polarizations of the beams to transverse polarizations. 
These developments have led to a series of investigations on
the use of transverse polarization in achieving these aims, see, 
e.g.~\cite{rizzo}.

One sensitive window to the possibility of observing new physics is
through the observation of CP violation in processes where it is
expected to be either absent of suppressed in the standard model.
In the context of CP violation,
the role of transverse polarization has been
studied in \cite{Burgess:1990ba,Choi:2001ww,Ananthanarayan:2003wi,
Ananthanarayan:2004eb,Rindani:2004ue}, whereas that of longitudinal polarization
in \cite{Ananthanarayan:2002fh,Rindani:2003av}, and in references quoted 
therein. The potential of longitudinal polarization to improve the sensitivity 
of CP-violating observables has been known for a long time. 
The transverse polarization potential
at the LC was recently proposed in the context of \ttbar 
production \cite{Ananthanarayan:2003wi}, where the need for chirality violating 
interactions for the observation of CP violation through top azimuthal 
distribution was emphasized. In case of a neutral final state, however, CP 
violation is possible to observe even with chirality conserving interactions. 
In $\gamma Z$ production
a CP-violating contribution can arise if anomalous CP-violating \ggz and \gzz 
couplings are present \cite{Czyz:1988yt,Choudhury:1994nt}.
The interference of the contributions from these
anomalous couplings with the SM contribution give rise to the 
polar-angle forward-backward asymmetry with unpolarized \cite{Czyz:1988yt} or 
longitudinally polarized beams \cite{Choudhury:1994nt}, as well as new 
combinations of polar
and azimuthal asymmetries in the presence of transversely polarized beams 
\cite{Ananthanarayan:2004eb}.

However, there  may be sources different
from anomalous triple-gauge-boson vertices that could also contribute
to such asymmetries.  A set of model-independent form factors that are
gauge invariant and chirality conserving were proposed as such sources
in ref.\cite{Abraham:1993zh} in the context of
$Z\to b\overline{b}\gamma$ events. It is the purpose of this work to make 
use of such general form factors for the process \eegz to examine CP-violating 
asymmetries in the presence of longitudinal or  transverse polarization. Our 
emphasis will however be on transverse polarization, since it provides a handle
on a different and larger set of form factors, as will be seen below.  
We employ these form factors and evaluate their
contribution to the differential cross section
and pertinent asymmetries to leading order.

 In general, these form factors
can be functions of both $s$ and $t$; here we consider the dependence  on $t$ to
be absent and treat them as constants at a fixed 
$\sqrt{s}$. The analysis would be considerably more complicated if we put in 
the 
dependence of form factors on $t$ as well. 

Closely related sources of CP violation have been constrained 
experimentally at the LEP collider \cite{Buskulic:1996ig} in the reaction
$Z\to b \overline{b}g$ (see also \cite{Bernreuther:1994vg})
and have been considered elsewhere \cite{Bernreuther:1996fi,Nachtmann:2003hg}.

In Sec.\ref{sec_process} we describe the form factors for the process of interest and compute
the differential cross section due to the SM and the anomalous
couplings, the latter to leading order.  In Sec.\ref{asymmetries} we
describe the construction of CP-odd asymmetries from which we can
extract the anomalous couplings and provide a detailed
discussion on their utility, followed by numerical results
in Sec.\ref{numerical}.  We find that the different anomalous couplings
can be constrained at a realistic LC with design luminosities
of 500 ${\rm fb}^{-1}$ at varying levels, lying between $10^{-4}-10^{-2}$.
In Sec.\ref{conclusions} we summarize our
conclusions.

\section{\boldmath The process \eegz with anomalous form \mbox{factors}}\label{sec_process}
The process considered is 
\begin{equation}
e^-(p_-,s_-)+e^+(p_+,s_+)\rightarrow \gamma (k_1,\alpha)+Z(k_2,\beta).
        \label{process}
\end{equation}
We shall assume that the amplitudes are generated by the standard model
as well as a general set of CP-violating interactions of the type
proposed by Abraham and Lampe \cite{Abraham:1993zh}.  They are completely 
determined by
vertex factors that we denote by $\Gamma^{SM}_{\alpha\beta}$
and $\Gamma_{\alpha\beta}$.  The vertex factor  corresponding to SM is given by
\begin{equation}
\Gamma^{SM}_{\alpha\beta}={e^2\over 4 \sin \theta_W \cos \theta_W}
 \left\{\gamma_\beta 
(g_V-g_A\gamma_5)
\frac{1}{\slp_- - \slk_1} 
\gamma_\alpha+
\gamma_\alpha 
\frac{1}{\slp_- -\slk_2}                
\gamma_\beta 
(g_V-g_A \gamma_5) 
\right\}.
\end{equation} 
In the above,
the vector and axial vector $Z$ couplings of the electron are 
\begin{equation}
 g_V= -1 + 4\sin^2\theta_W ;\quad g_A = -1
     \label{gVgA}.
\end{equation}
The anomalous form factors may be introduced via the following
vertex factor:
\begin{eqnarray}\label{anom}
& \displaystyle \Gamma_{\alpha\beta}={i e^2 \over 4 \sin\theta_W \cos\theta_W}
\left\{\frac{1}{m_Z^4}\left((v_1+  a_1 \gamma_5)\gamma_\beta(
2 p{_-}_\alpha (p_+\cdot k_1)-
2 p{_+}_\alpha (p_-\cdot k_1)) + \right. \right. & \nonumber \\ 
& \displaystyle \left. \left.
 ((v_2+  a_2 \gamma_5) p{_-}_\beta + 
(v_3+  a_3 \gamma_5) p{_+}_\beta)(\gamma_\alpha 2 p_-\cdot k_1-
2 p{_-}_\alpha \slk_1)+
\right. \right.& \nonumber \\
& \displaystyle \left. \left.
 ((v_4+  a_4 \gamma_5) p{_-}_\beta +
(v_5+  a_5 \gamma_5) p{_+}_\beta)(\gamma_\alpha 2 p_+\cdot k_1-
2 p{_+}_\alpha \slk_1)\right)+ \right. & \nonumber \\
& \displaystyle \left. \frac{1}{m_Z^2} 
(v_6+  a_6 \gamma_5)(\gamma_\alpha k_{1\beta}-\slk_1 
g_{\alpha\beta}) 
 \right\} .&
\end{eqnarray}
This is the most general form of coupling consistent with Lorentz invariance, gauge invariance and 
chirality conservation. These couplings include  contact interactions, as well as contributions from triple gauge vertices considered in \cite{ Choudhury:1994nt,Ananthanarayan:2004eb}. The latter would be a special case of our general interactions.
We note here that  not all the form factors are CP violating. 
The following combinations are CP odd:
$r_2+r_5,\,  r_3+r_4,\,  r_6;\, r=v,a$.
The combinations $r_1,  \,r_2-r_5, \, r_3-r_4;\, r=v,a$,  
are even under CP.

When the $e^-$ and $e^+$ beams have longitudinal polarizations $P_L$ 
and $\overline{P}_L$, 
we obtain the differential cross section
for the process (1) to be
\be
\displaystyle
\frac{d\sigma}{d\Omega    }_L =
    {\cal B}_L\left(1-P_L\overline{P}_L\right)
\left[
       \frac{1}{\sin^2 \theta}
          \left( 1 + \cos^2 \theta + \frac{4 \sbar}{( \sbar - 1)^2}
           \right)
     + C_L              
\right]  \; ,
    \label{diff c.s.L}
\ee
where
\begin{eqnarray}
& \displaystyle
\sbar  \equiv  \frac{s}{m_Z^2},\,\,
   {\cal B}_L  = \frac{\alpha^2}{16 \sin^2\theta_W m_W^2 \sbar}
     \left( 1 - \frac{1}{\sbar}   \right)
     (g_V^2+g_A^2-2Pg_Vg_A) , & 
\end{eqnarray}
with
\be
P = \frac{P_L - \overline{P}_L}{1- P_L \overline{P}_L},
\ee
and
 \begin{eqnarray}
& \displaystyle
C_{L}  =  
        \frac{1}{4 (g_V^2+g_A^2-2Pg_Vg_A)}
    \left\{\sum_{i=1}^6
\left(	(g_V-Pg_A) {\rm Im}v_i+ (g_A-Pg_V) {\rm Im}a_i\right) X_i 
\right\}.&
\end{eqnarray}
The differential cross section for transverse polarizations $P_T$ and $\overline{P}_T$ of $e^-$ and $e^+$ is given by
\be
\displaystyle
\frac{d\sigma}{d\Omega    } =
    {\cal B}_T
\left[
       \frac{1}{\sin^2 \theta}
          \left( 1 + \cos^2 \theta + \frac{4 \sbar}{( \sbar - 1)^2}
	- P_T \overline{P}_T\frac{g_V^2-g_A^2}{g_V^2+g_A^2}
	\sin^2 \theta \cos 2\phi 
           \right)
     + C_T              
\right]  \; ,
    \label{diff c.s.T}
\ee
where $\bar{s}$ is as before, 
\begin{eqnarray}
& \displaystyle
   {\cal B}_T  = \frac{\alpha^2}{16 \sin^2\theta_W m_W^2 \sbar}
     \left( 1 - \frac{1}{\sbar}   \right)
     (g_V^2+g_A^2) , & 
\end{eqnarray}
and
 \begin{eqnarray}
& \displaystyle
C_{T}  =
        \frac{1}{4 (g_V^2+g_A^2)}
    \left\{\sum_{i=1}^6
        (g_V {\rm Im}v_i+ g_A {\rm Im}a_i) X_i +\right. & \nonumber \\
        & \displaystyle
    \left. P_T \overline{P}_T\,
\sum_{i=1}^6\left(
        (g_V {\rm Im}v_i- g_A {\rm Im}a_i)
    \cos 2\phi+
        (g_A {\rm Re}v_i- g_V {\rm Re}a_i)
                 \sin 2\phi\right)Y_i  \right\} &
   \label{notation}
\end{eqnarray}
$X_i,\, Y_i \, (i=1,...6)$ are given in Table 1.
\begin{table}\label{xytable}
\begin{center}
\begin{tabular}{||c|c|c||}\hline
$i$ &  $X_i$ & $Y_i$ \\ \hline \hline
$1$ & $-2  \sbar (\sbar+1)$ & 0  \\ \hline
$2$ & $  \sbar (\sbar-1) (\cos\theta-1)$ & $0$\\ \hline
$3$ & $  0 $ & $\sbar(\sbar-1)(\cos\theta-1)$ \\ \hline
$4$ & $  0 $ & $\sbar(\sbar-1)(\cos\theta+1)$ \\ \hline
$5$ & $ \sbar(\sbar-1) (\cos\theta+1)$ & $0$ \\ \hline
$6$ & $2  (\sbar-1) \cos\theta $ & $2 (\sbar-1)\cos\theta$\\ \hline
\end{tabular}
\caption{The contribution of the new couplings to the polarization
independent and dependent parts of the cross section}
\end{center}
\end{table}
In the expressions above
$\theta$ is the angle between photon and the $e^-$ directions, and $\phi$
is the azimuthal angle of the photon, with $e^-$ direction chosen as the $z$
axis and the direction of its transverse polarization chosen as the $x$ axis.
The $e^+$ transverse polarization direction is chosen parallel to the 
$e^-$ transverse polarization
direction. 

We have kept only terms of leading order in the anomalous couplings, 
since they are expected to be small. The above expression may be 
obtained either 
by using standard trace techniques for Dirac spinors with a
transverse spin four-vector, 
or by first calculating helicity amplitudes and then writing
transverse polarization states in terms of 
helicity states. 
We note that the contribution of the interference between 
the SM amplitude and the anomalous amplitude vanishes for $s=m_Z^2$. The reason
for this is that for $s=m_Z^2$ the photon in the final state is produced with
zero energy and momentum, and for for the photon four-momentum $k_1=0$, the anomalous contribution  (\ref{anom}) vanishes identically.
A noteworthy feature of the result is that with the exception
of the case of $v_6$ and $a_6$, the anomalous form factors either contribute
to the transverse polarization dependent part, or to the 
longitudinal polarization dependent and polarization independent parts
of the differential cross section, but not both.  
It is only for the case of
$i=6$ that the differential cross section receives contribution to
both.  We note that the results corresponding to the
case of the anomalous triple-gauge-boson vertices
\cite{Ananthanarayan:2004eb} is reproduced by the choice $v_i=a_i=0\,
(i\ne 6)$, 
$v_6 =  ( g_V \lambda_1 - \lambda_2)/2$ and $a_6 =  g_A \lambda_1/2$.

It is also interesting to note that the combination $r_2+r_5 + r_3 + r_4$ give the same angular distribution as $r_6$, and that the combination $r_2 - r_5$ gives the same angular distribution as $r_1$
(with $r$ standing for $v$ and $a$ in both cases). This implies that so far as the angular distribution from the interference terms is concerned, the number of independent form factors is less than what is displayed in eq. (\ref{anom}). In fact, there are only 6 independent quantities that can be determined by the angular distribution, which are the coefficients of the various combinations of trigonometric functions occurring in the angular distribution, of which 3 are CP violating. On the other hand, the number of independent form factors being 12, the number of real parameters it corresponds to is 24. Clearly, not all these can be determined by the angular distribution, but only certain linear combinations. 
Moreover,  so far as the real parts of form factors are concerned, 
it is only the  combinations $g_A{\rm Re}v_i -g_V {\rm Re}a_i$ which appear. 
Thus it is not possible to separately determine the real parts of $v_i$ 
and $a_i$.

The angular distribution derived above can be used to construct various asymmetries which can isolate CP-conserving as well as CP-violating combinations of form factors. We will however concentrate only on CP-violating form factor in what follows.

\section{CP-odd asymmetries}\label{asymmetries}

We now present a discussion of the possible CP-odd asymmetries in the process. 

We first take up the case of transverse polarization.
In order to understand the CP properties of various terms in the differential
cross section, we note the following relations:
\beq\label{ctheta}
\vec{P}\cdot \vec{k}_1 =\frac{\sqrt{s}}{2} \vert \vec{k}_1 \vert \cos\theta\;,
\eeq
\beq\label{s2ts2p}
(\vec{P} \times \vec{s}_- \cdot \vec{k}_1)( \vec{s}_+\cdot \vec{k}_1) + 
(\vec{P} \times \vec{s}_+ \cdot \vec{k}_1) (\vec{s}_-\cdot \vec{k}_1) 
= \frac{\sqrt{s}}{2} \vert \vec{k}_1 \vert^2 \sin^2\theta \sin 2\phi\; ,
\eeq
\beq\label{s2tc2p}
(\vec{s}_- \cdot \vec{s}_+) (\vec{P}\cdot \vec{P} \vec{k}_1 \cdot \vec{k}_1 - 
\vec{P}\cdot\vec{k}_1 \vec{P}\cdot\vec{k}_1) - 2 (\vec{P}\cdot \vec{P}) 
( \vec{s}_-
\cdot \vec{k}_1) ( \vec{s}_+\cdot \vec{k_1})
 = -\frac{s}{4} \vert \vec{k}_1 \vert^2 
\sin^2\theta \cos 2\phi\; ,
\eeq 
where $\vec{P}=\frac{1}{2}(\vec{p}_- - \vec{p}_+)$, and it is assumed that $\vec{s}_+=\vec{s}_-$.
Observing that the vector $\vec{P}$ is C and P odd, that the photon
momentum $\vec{k}_1$ is C even but P odd, and that the spin vectors
$\vec{s}_{\pm}$ are P even, and go into each other under C, 
we can immediately check that only the left-hand side (lhs)
of eq. (\ref{ctheta}) is CP odd, while the lhs of 
eqs. (\ref{s2ts2p}) and (\ref{s2tc2p})
are CP even. Of all the above, only the lhs of (\ref{s2ts2p}) is odd under
naive time reversal T. 
In the light of the observations above, as well as the general
discussion provided in the previous section on the CP properties
of (combinations of) the form factors, we note that it is only
the coefficients of 
$r_2+r_5,\,  r_3+r_4,\,  r_6,\, r=v,a$ that have a pure $\cos\theta$
dependence.  Consequently, the coefficients of the 
combinations $r_1,  \,r_2-r_5, \, r_3-r_4,\, r=v,a$,  
have no $\cos\theta$ dependence. Moreover, invariance under CPT implies that terms with the right-hand side (rhs) of (\ref{ctheta}) by itself, or multiplying the rhs of (\ref{s2tc2p}) would occur with absorptive (imaginary) parts of the form factors, whereas the rhs of (\ref{ctheta}) multiplied by the rhs of (\ref{s2ts2p}) would appear with dispersive (real) parts of the form factors. We will see this explicitly below when we construct asymmetries which isolate the various angular dependences.
 
For longitudinal polarization, in addition to (\ref{ctheta}), there is 
another CP-odd quantity, viz., 
\beq\label{cthetapol}
\frac{1}{2}\left(\vec{s}_- + \vec{s}_+\right) \cdot \vec{k}_1 
= \vert \vec{k}_1 \vert \cos\theta\;.
\eeq
While this is also proportional to $\cos\theta$ like (\ref{ctheta}), it is 
expected to appear with a factor $(P_L-\overline{P}_L)$ multiplying it. It is 
also CPT odd, and would therefore occur with the absorptive parts of form 
factors.

We now proceed to construct asymmetries of interest and derive
the numerical consequences to the anomalous form factors.
We begin by noting that we shall
assume a cut-off  $\thmin$ on the polar angle
$\theta$ of the photon
in the forward and backward directions. This cut-off is needed to stay away
from the beam  pipe. It can further be chosen to optimize the sensitivity.
The total cross section corresponding to the cut
$\thmin < \theta < \pi - \thmin$
can then be easily obtained by integrating the differential cross section
above.

We now define the following CP-odd asymmetries, 
$A_1(\theta_0),\, A_2(\theta_0), A_3(\theta_0)$\footnote{Alternatively,
with transverse polarization we could use for $A_3$ the definition 
of ref.\cite{Ananthanarayan:2004eb} which
would receive contributions from both polarization
independent and dependent parts of the cross sections.  
This would then result in
$
A_3(\theta_0)=        {\cal B}'_T \,
 \frac{ \pi}{2}\left[  g_A\left\{\sbar ({\rm Im} a_2+{\rm Im} a_5)+
2 {\rm Im} a_6   \right\}+     g_V \left\{\sbar ({\rm Im} v_2+{\rm Im} v_5) +2{\rm Im} v_6 \right\}                       				\right]
      +
A_2(\theta_0).$
With polarization flips, it would then be possible to
separate real and imaginary parts of $r_3+r_4+2r_6/\sbar$, $( r=v,a)$, 
from $A_1$ and
$A_2,$ and imaginary parts of $r_2+r_5+2 r_6/\sbar$,  $(r=v,a)$ from this $A_3$.
With the present definition, however, the role of longitudinal polarization in 
enhancing the sensitivity of observables is particularly transparent.}
which combine, in general, a
forward-backward asymmetry with an appropriate asymmetry in $\phi$, so as to
isolate appropriate anomalous couplings:
\begin{eqnarray}& \displaystyle A_1=
{1\over \sigma_0}
\sum_{n=0}^3 (-1)^n
\left(
\int_{0}^{\cos \theta_0} d \cos\theta
 -
\int_{-\cos \theta_0}^{0} d \cos\theta \right) 
 \int_{\pi n/ 2}^{\pi(n+1)/  2} d\phi \,
{d \sigma \over d \Omega} , 
\end{eqnarray}
\begin{eqnarray}& \displaystyle A_2=
{1\over \sigma_0}
\sum_{n=0}^3(-1)^n \left(
\int_{0}^{\cos \theta_0} d \cos\theta -
\int_{-\cos \theta_0}^0 d \cos\theta \right)
 \int_{\pi (2 n-1)/4}^{\pi(2 n+1)/4} d\phi \,
{d \sigma \over  d \Omega} , &
\end{eqnarray}
and
\begin{eqnarray}
& \displaystyle A_3(\thmin)=\frac{1}{\sigma_0}
\left(\int_{-\cos \theta_0}^{0} d \cos\theta-
\int^{\cos \theta_0}_0 d \cos\theta\right)
\int_{0}^{2 \pi} d\phi \,
{d \sigma \over d \Omega            }\, , &
\end{eqnarray}
with
\begin{eqnarray}
& \displaystyle \sigma_0 \equiv \sigma_0(\thmin)=
\int_{-\cos \theta_0}^{\cos \theta_0} d \cos\theta
\int_{0}^{2 \pi} d\phi \,
{d \sigma \over d \Omega            }\, . &
\end{eqnarray}
Of the asymmetries above, $A_1$ and $A_2$ exist only in the presence of 
transverse polarization, and are easily evaluated to be
\begin{equation}
 \displaystyle A_1(\theta_0)= 
\displaystyle               
         {\cal B}'_T       
	\,  P_T \overline{P}_T\, \left[
g_A \left\{\sbar 
({\rm Re} v_3+{\rm Re}v_4)+2 {\rm Re} v_6\right\}- 
g_V \left\{\sbar 
({\rm Re} a_3+{\rm Re}a_4)+2 {\rm Re} a_6\right\}\right], 
		 \; 
		 \end{equation}
\begin{equation}
 \displaystyle A_2(\theta_0)= 
\displaystyle               
         {\cal B}'_T  
	\,  P_T \overline{P}_T\, \left[
g_V \left\{\sbar 
({\rm Im} v_3+{\rm Im}v_4)+2 {\rm Im} v_6\right\}- 
g_A\left\{\sbar 
({\rm Im} a_3+{\rm Im}a_4)+2 {\rm Im} a_6\right\}\right], 
		 \; 
		 \end{equation}
In the equations above, we have defined
\begin{equation}
{\cal B}'_T =
\frac{{\cal B}_T (\sbar - 1)\cos^2\theta_0}  
	{ (g_V^2+g_A^2)\sigma_0^T}.
\end{equation}
with
\begin{eqnarray}
& \displaystyle \sigma_0^T=4 \pi {\cal B}_T
  \left[ \left\{ \frac{\sbar^2 + 1}{(\sbar - 1)^2}
                   \ln \left( \frac{1 + \cmin }
                                   {1 - \cmin }
                            \right)
                 - \cmin
          \right\}\right]. \; 
\end{eqnarray}

The asymmetry $A_3$ is independent of transverse polarization and is found to be
\begin{eqnarray}
\displaystyle A_3(\theta_0)&= 
 \displaystyle
       \; {\cal B}'_L 
 \frac{ \pi}{2}&\!\left[  (g_A-Pg_V)\left\{\sbar ({\rm Im} a_2+{\rm Im} a_5)+
2 {\rm Im} a_6   \right\} \right. \nonumber \\
& \;\;  & \left.\! \!+ (g_V-Pg_A)
 \left\{\sbar ({\rm Im} v_2+{\rm Im} v_5) +2{\rm Im} v_6 \right\}                       				\right] ,
    \end{eqnarray}
where
\begin{equation}
{\cal B}'_L =
\frac{{\cal B}_L (1-P_L\overline{P}_L)(\sbar - 1)\cos^2\theta_0}
        { (g_V^2+g_A^2-2Pg_Vg_A)\sigma_0^L}.
\end{equation}
with
\begin{eqnarray}
& \displaystyle \sigma_0^L=4 \pi {\cal B}_L(1-P_L\overline{P}_L)
  \left[ \left\{ \frac{\sbar^2 + 1}{(\sbar - 1)^2}
                   \ln \left( \frac{1 + \cmin }
                                   {1 - \cmin }
                            \right)
                 - \cmin
          \right\}\right]. \;
\end{eqnarray}

We now make some observations on the above expressions which justify the 
choice of our asymmetries and highlight the novel features of our work.
It can be seen that
$\displaystyle A_1(\theta_0)$ is proportional to combinations of
${\rm Re} v_i$, ${\rm Re} a_i$,
and the other two asymmetries depend on combinations of
${\rm Im} v_i$, ${\rm Im} a_i$.
Indeed, but for the case $i=6$, one of the latter asymmetries depends
on a specific combination of couplings that is complementary to that
which shows up in the other. The case of the anomalous triple-gauge-boson vertex is
similar to that of the case $i=6$ since in this case there are  contributions to both
the polarization dependent as well as the polarization independent part
of the cross section.

\section{Numerical Results}\label{numerical}

We have several form factors, and if all of them are present simultaneously, the analysis of numerical results would be complicated. We therefore choose one form factor to be nonzero at a time to discuss numerical results.  

\begin{figure}[t]
\centering
\vskip -1.5cm
\psfig{file=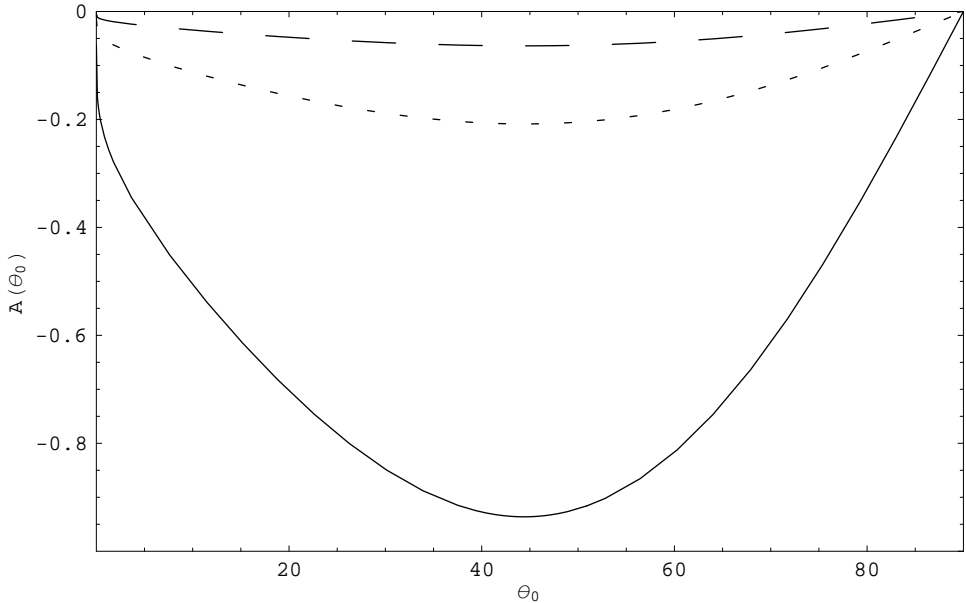,height=10cm}
\caption{The asymmetries $A_1(\theta_0)$ (solid line), $A_2(\theta_0)$ (dashed line) and $A_3(\theta_0)$ (dotted line), defined in the text, plotted as functions
of the cut-off $\theta_0$ for a value of Re $v_6 ={\rm Im~}v_6 = 1$.}
\label{asym1}
\end{figure}

We first take up for illustration the case when only  ${\rm Re}~v_6$ is nonzero,
 since the results for other CP-violating combinations can be deduced from this 
case. 
We choose $P_T=0.8$ and $\overline{P}_T=0.6$, and vanishing longitudinal 
polarization for this case. 
Fig. \ref{asym1} shows the asymmetries $A_i$ as a 
function of the cut-off 
when the values of the anomalous couplings ${\rm Re}~v_6$ (for the case of 
$A_1$) and Im~$v_6$ (for the case of $A_2$ and $A_3$) alone
are set to unity.
The asymmetries vanish not only for $\theta_0=0$, by definition, but also
for $\theta_0=90^{\circ}$, because they are proportional to $\cos\theta_0$.  
Also, they peak  at around $45^{\circ}$. 

We have calculated 90\% CL limits that can be obtained with a 
LC with
$\sqrt{s} = 500$ GeV, $\int L dt = 500$ fb$^{-1}$, $P_T = 0.8$, and
$\overline{P}_T = 0.6$ making use of the asymmetries $A_i$ ($i=1,2$).
For $A_3$, we assume unpolarized beams.

The limiting value $v^{\rm lim}$ (\ie the respective real or imaginary
part of the coupling) is related to the value $A$ of the asymmetry for unit
value of the coupling constant.
\beq
v^{\rm lim} = \frac{1.64}{|A|\sqrt{N_{SM}}},
\eeq
where $N_{SM}$ is the number of SM events.
\begin{figure}[htb]
\centering
\vskip 1.5cm
\psfig{file=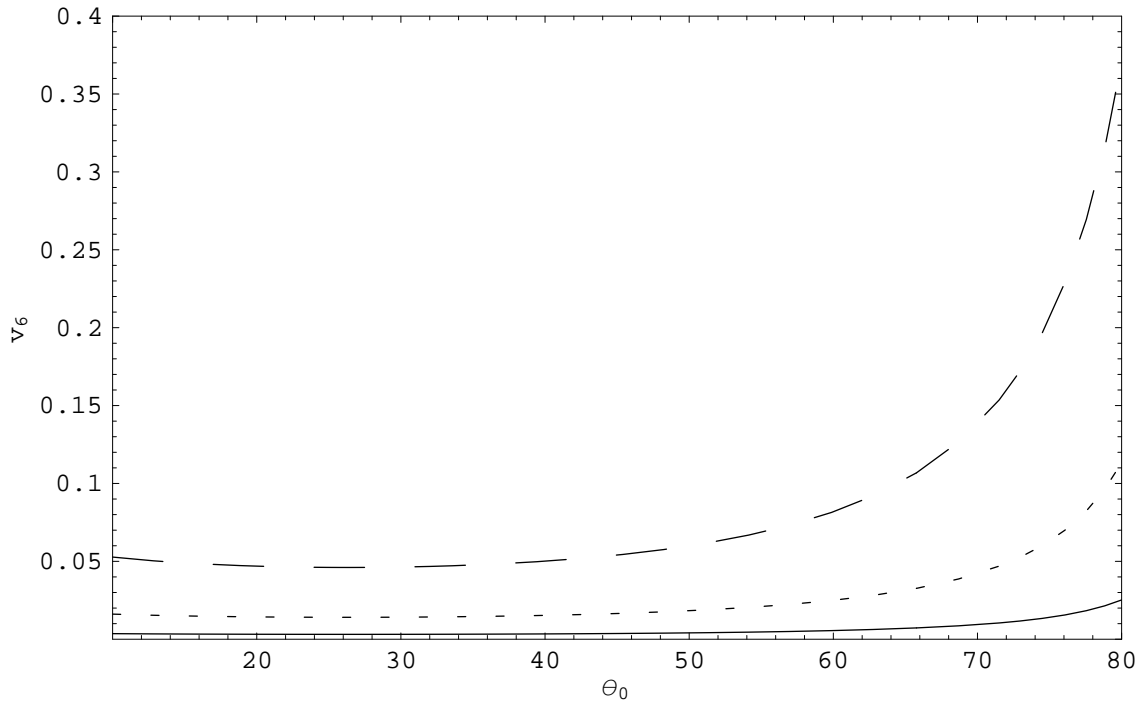}
\caption{The 90\% C.L. limit on Re~$v_6$ 
from the asymmetry $A_1(\theta_0)$ (solid line), and on Im~$v_6$ from $A_2(\theta_0)$ (dashed line) and  $A_3(\theta_0)$ (dotted line),
plotted as functions of the cut-off $\theta_0$.}
\label{a1lim}
\end{figure}

The curves from $A_1$ corresponding to setting only Re~$v_6$ nonzero, and from $A_2$ and $A_3$ corresponding to keeping only Im~$v_6$ nonzero are illustrated in Fig. \ref{a1lim}.
We note that there is a stable plateau for a choice of
$\theta_0$ such that $10^{\circ} \lsim \theta_0 \lsim 40^{\circ}$; and we choose the optimal value of
$26^0$.  The sensitivity corresponding to this for 
Re $v_6$ is $\sim 3.1\cdot 10^{-3}$.
 
The results for the other couplings may be
inferred in a straightforward manner from the explicit example above.
For the asymmetry $A_1$, if we were to set $v_3(v_4)$ to unity,
with all the other couplings to zero, then the asymmetry would be
simply scaled up by a value $\sbar/2$, which for the case at hand is
$\simeq 14.8$.  The corresponding limiting value would
be suppressed by the reciprocal of this factor.

The results for the couplings ${\rm Re}~a_i,\, i=2,5,6$, compared to
what we have for the vector couplings 
would be scaled by a factor $g_V/g_A\simeq 0.07$ for the asymmetries
and by the reciprocal of this factor for the sensitivities.

The results coming out of the asymmetry $A_2$ are such that the 
sensitivities of the imaginary parts of
$v$ and $a$ are interchanged {\it vis \`a vis} what
we have for the real parts coming out of $A_1$.

The final set of results we have is for the form factors that
may be analyzed via the asymmetry $A_3$,  
 which depends only on longitudinal polarizations. We treat the cases
of unpolarized beams and longitudinally polarized beams with $P_L = 0.8$, and
$\overline{P}_L = -0.6$ separately.
For the unpolarized case,
the results here for Im $v_6$
correspond to those coming from $A_2$, with the asymmetry scaled
up now by a factor corresponding to $\pi/2$ and a further factor
$(P_T \overline{P}_T)^{-1}$ $(\simeq~2.1)$, which yields an overall factor
of $\sim 3.3$.  The corresponding sensitivity is smaller is by
the same factor.  Indeed, the results we now obtain for ${\rm Im}v_i,\, i=3,4$
are related to those obtained from $A_2$ for $i=2,5$ by the same
factor.

For the case with longitudinal polarization, the sensitivities for the 
relevant ${\rm Im}v_i$ are enhanced by almost an order of magnitude, whereas the
sensitivities for ${\rm Im}a_i$ are improved marginally. For the case of 
anomalous triple gauge-boson couplings contributing to the process, a similar 
conclusion was obtained in \cite{Choudhury:1994nt}.

All the results discussed above are now summarized in the 
Tables 2, 3 and 4. 
\begin{table}\label{summ1}
\begin{center}
\begin{tabular}{||c|c|c||c|c|c||}\hline\hline
\multicolumn{3}{||c||} {$A_1$} &
\multicolumn{3}{||c||} {$A_2$}  
\\ \hline \hline
Re $v_3$ & Re $v_4$ & Re $v_6$ & 
Im $v_3$ & Im $v_4$ & Im $v_6$  \\ \hline
$2.1 \cdot 10^{-4}$ & $2.1 \cdot 10^{-4}$ & $3.1 \cdot 10^{-3}$ &
$3.1 \cdot 10^{-3}$ & $3.1 \cdot 10^{-3}$ & $4.6 \cdot 10^{-2}$\\ \hline \hline
Re $a_3$ & Re $a_4$ & Re $a_6$ & 
Im $a_3$ & Im $a_3$ & Im $a_6$   \\ \hline
$3.1 \cdot 10^{-3}$ & $3.1 \cdot 10^{-3}$ & $4.6 \cdot 10^{-2}$ &
$2.1 \cdot 10^{-4}$ & $2.1 \cdot 10^{-4}$ & $3.1 \cdot 10^{-3}$ \\ \hline\hline
 \end{tabular}
\caption{Table of sensitivities obtainable at the LC with the
machine and operating parameters given in the text for
the asymmetries $A_1$ and $A_2$.}
\end{center}
\end{table}

\begin{table}\label{summ2}
\begin{center}
\begin{tabular}{||c|c|c||}\hline\hline
\multicolumn{3}{||c||} {$A_3$} \\ \hline \hline
Im $v_2$ & Im $v_5$ & Im $v_6$ \\ \hline
$9.3 \cdot 10^{-4}$ & $9.3 \cdot 10^{-4}$ & $1.4 \cdot 10^{-2}$ \\
\hline \hline
Im $a_2$ & Im $a_5$ & Im $a_6$ \\ \hline
$6.4 \cdot 10^{-5}$ & $6.4 \cdot 10^{-5}$ & $9.6 \cdot 10^{-4}$ \\ \hline\hline
\end{tabular}
\caption{Table of sensitivities obtainable at the LC with the
machine and operating parameters given in the text for
the asymmetries $A_3$ with unpolarized or transversely polarized beams.
}
\end{center}
\end{table}

\begin{table}\label{summ3}
\begin{center}
\begin{tabular}{||c|c|c||}\hline\hline
\multicolumn{3}{||c||} {$A_3$} \\ \hline \hline
Im $v_2$ & Im $v_5$ & Im $v_6$ \\ \hline
$5.6 \cdot 10^{-5}$ & $5.6 \cdot 10^{-5}$ & $8.4 \cdot 10^{-4}$ \\
\hline \hline
Im $a_2$ & Im $a_5$ & Im $a_6$ \\ \hline
$5.2 \cdot 10^{-5}$ & $5.2 \cdot 10^{-5}$ & $7.9 \cdot 10^{-4}$ \\ \hline\hline
\end{tabular}
\caption{Table of sensitivities obtainable at the LC with the
machine and operating parameters given in the text for
the asymmetries $A_3$ with longitudinally polarized beams.
}
\end{center}
\end{table}

\newpage
\section{Conclusions}\label{conclusions}
Forward-backward asymmetry of a neutral particle with 
polarized beams as a signal of CP violation has
been studied here in some generality. 
We have considered a general form-factor parametrization
and have isolated from these (combinations of) CP-violating
form factors.  Only one out of these corresponding to
$i=6$ has the special property
of contributing to both polarization dependent as well as
independent parts of the cross section.  Two out of the rest
corresponding to $i=2,5$ can have observable consequences
in the absence of transverse polarization, 
while those 
corresponding to $i=3,4$ can only be studied in the
presence of transverse polarization.
Since the former ones occur in the asymmetry $A_3$ which is even under naive time reversal, 
the CPT theorem implies that in such a case the
asymmetry is proportional to the absorptive part of the amplitude. 
The sensitivities for Im~$v_i$ ($i=2,5$) are improved by an order of magnitude 
with the use of longitudinal polarization, whereas the sensitivities 
for Im~$a_i$ ($i=2,5$) are improved only marginally.
The asymmetry
$A_1$ that we study in the presence of transverse polarizations 
includes also an
azimuthal angle asymmetry, which makes it odd under naive time reversal. It is
thus proportional to the real part of the couplings. This real part
cannot be studied without transverse polarization. 

In general, one can conclude that longitudinal beam polarization plays a useful 
role in 
improving the sensitivity to absorptive parts of CP-violating form factors, 
which 
are amenable to measurement even without polarization. However, transverse 
polarization enables measurement of dispersive parts of certain form factors 
which are inaccessible without polarization or with longitudinal polarization.

This work extends recent results where CP violation due to
anomalous triple-gauge-boson vertices was considered.
Anomalous triple-gauge-boson couplings would occur at loop level through triangle diagrams in theories like minimal supersymmetric standard model (MSSM) \cite{Choudhury:2000bw} or multi-Higgs models involving particles beyond SM coupling to gauge bosons.  The form factors we consider here include these contributions, as well as additional form factors which might also arise in these theories through box diagrams \cite{Gounaris:2002za}. It is thus natural to include all form factors. 
We have shown that with typical LC energies and realistic
integrated luminosities and degrees of electron and positron
beam polarization, a window of opportunity for the discovery
of new physics can be opened.

\vskip .2cm
\noindent Acknowledgements: We thank Ritesh K. Singh for collaboration
at the initial stages of this work. We also thank the organizers of WHEPP8 (8th Workshop on High Energy Physics Phenomenology), held at the  Indian Institute of Technology, Mumbai, during January 5-16, 2004, for hospitality and a stimulating atmosphere, where the idea for this work originated. BA thanks the
Department of Science and Technology, Government of India, 
and the Council for Scientific and Industrial Research, Government
of India for support. SDR acknowledges financial assistance under the COE Fellowship Program and thanks Yukinari Sumino for hospitality at the High Energy Theory Group, Tohoku University.


\newpage
\begin{thebibliography}{99}
\bibitem{rizzo}
T.~G.~Rizzo,
JHEP {\bf 0302} (2003) 008
[arXiv:hep-ph/0211374];
T.~G.~Rizzo,
JHEP {\bf 0308} (2003) 051
[arXiv:hep-ph/0306283], and references therein.



\bibitem{Burgess:1990ba}
C.~P.~Burgess and J.~A.~Robinson,
Int.\ J.\ Mod.\ Phys.\ A {\bf 6} (1991) 2707.

\bibitem{Choi:2001ww}
S.~Y.~Choi, J.~Kalinowski, G.~Moortgat-Pick and P.~M.~Zerwas,
Eur.\ Phys.\ J.\ C {\bf 22} (2001) 563
[Addendum-ibid.\ C {\bf 23} (2002) 769]
[arXiv:hep-ph/0108117].

\bibitem{Ananthanarayan:2003wi}
B.~Ananthanarayan and S.~D.~Rindani,
Phys.\ Rev.\ D {\bf 70} (2004) 036005
[arXiv:hep-ph/0309260];
LC Note LC-TH-2003-099.

\bibitem{Ananthanarayan:2004eb}
B.~Ananthanarayan, S.~D.~Rindani, R.~K.~Singh and A.~Bartl,
Phys.\ Lett.\ B {\bf 593}, 95 (2004)
[arXiv:hep-ph/0404106].

\bibitem{Rindani:2004ue}
S.~D.~Rindani,
Phys.\ Lett.\ B {\bf 602} (2004) 97
[arXiv:hep-ph/0408083].

\bibitem{Ananthanarayan:2002fh}
B.~Ananthanarayan, S.~D.~Rindani and A.~Stahl,
Eur.\ Phys.\ J.\ C {\bf 27}, 33 (2003)
[arXiv:hep-ph/0204233].

\bibitem{Rindani:2003av}
S.~D.~Rindani,
Pramana {\bf 61}, 33 (2003)
[arXiv:hep-ph/0304046].

\bibitem{Czyz:1988yt}
H.~Czyz, K.~Kolodziej and M.~Zralek,
Z.\ Phys.\ C {\bf 43} (1989) 97.

\bibitem{Choudhury:1994nt}
D.~Choudhury and S.~D.~Rindani,
Phys.\ Lett.\ B {\bf 335} (1994) 198
[arXiv:hep-ph/9405242].



\bibitem{Abraham:1993zh}
K.~J.~Abraham and B.~Lampe,
Phys.\ Lett.\ B {\bf 326}, 175 (1994).

\bibitem{Buskulic:1996ig}
D.~Buskulic {\it et al.}  [ALEPH Collaboration],
Phys.\ Lett.\ B {\bf 384}, 365 (1996).

\bibitem{Bernreuther:1994vg}
W.~Bernreuther, G.~W.~Botz, D.~Bruss, P.~Haberl and O.~Nachtmann,
Z.\ Phys.\ C {\bf 68}, 73 (1995)
[arXiv:hep-ph/9412268].

\bibitem{Bernreuther:1996fi}
W.~Bernreuther, A.~Brandenburg, P.~Haberl and O.~Nachtmann,
Phys.\ Lett.\ B {\bf 387}, 155 (1996)
[arXiv:hep-ph/9606379].

\bibitem{Nachtmann:2003hg}
O.~Nachtmann and C.~Schwanenberger,
Eur.\ Phys.\ J.\ C {\bf 32}, 253 (2004)
[arXiv:hep-ph/0308198].

\bibitem{Choudhury:2000bw}
D.~Choudhury, S.~Dutta, S.~Rakshit and S.~D.~Rindani,
Int.\ J.\ Mod.\ Phys.\ A {\bf 16}, 4891 (2001)
[arXiv:hep-ph/0011205].

\bibitem{Gounaris:2002za}
G.~J.~Gounaris, J.~Layssac and F.~M.~Renard,
Phys.\ Rev.\ D {\bf 67} (2003) 013012
[arXiv:hep-ph/0211327].





\end{thebibliography}
\end{document}